\documentclass[fleqn,usenatbib]{mnras}

\usepackage{graphicx}	
\usepackage{amsmath}	
\usepackage{amssymb}	
\usepackage{multirow}
\usepackage{makecell}

\newcommand{\kms}{\,km~s$^{-1}$} %
\def \deg{\,$^\circ$}%
\newcommand{\nb}{\,$N$-body}%
\newcommand{\merc}{\,\textsc{mercury}}%
\newcommand{\pc}{\,Pluto-Charon}%
\newcommand{\rp}{\,$r_{\textrm{P}}$}%
\newcommand{\gcm}{\,$\textrm{g cm}^{-3}$}%
\newcommand{\sols}{\,Solar System}%

\title{The Fate of Debris in the Pluto-Charon System}

\author[Smullen \& Kratter]{
Rachel~A.~Smullen,$^{1}$\thanks{E-mail: rsmullen@email.arizona.edu}
Kaitlin~M.~Kratter,$^{1}$
\\
$^{1}$Steward Observatory, University of Arizona, Tucson, AZ 85721, USA\\
}

\date{Accepted XXX. Received YYY; in original form ZZZ}

\pubyear{2016}

\begin{document}
\label{firstpage}
\pagerange{\pageref{firstpage}--\pageref{lastpage}}
\maketitle

\begin{abstract}

The Pluto-Charon system has come into sharper focus following the fly by of \emph{New Horizons}. We use $N$-body simulations to probe the unique dynamical history of this binary dwarf planet system.  We follow the evolution of the debris disc that might have formed during the Charon-forming giant impact. First, we note that in-situ formation of the four circumbinary moons is  extremely difficult if Charon undergoes eccentric tidal evolution.   We track collisions of disc debris with Charon, estimating that hundreds to hundreds of thousands of visible craters might arise from 0.3--5 km radius bodies. \emph{New Horizons} data suggesting a dearth of these small craters may place constraints on the disc properties. While tidal heating will erase some of the cratering history, both tidal and radiogenic heating may also make it possible to differentiate disc debris craters from Kuiper belt object craters. We also track the debris ejected from the Pluto-Charon system into the Solar System; while most of this debris is ultimately lost from the Solar System, a few tens of 10--30 km radius bodies could survive as a Pluto-Charon collisional family.  Most are plutinos in the 3:2 resonance with Neptune, while a small number populate nearby resonances.  We show that migration of the giant planets early in the Solar System's history would not destroy this collisional family. Finally, we suggest that identification of such a family would likely need to be based on composition as they show minimal clustering in relevant orbital parameters.

\end{abstract}




\section{Introduction}

\emph{New Horizon's} arrival at Pluto has brought a new spotlight to the \sols's largest Kuiper belt dwarf planet and most well-known binary.  Pluto and its largest moon, Charon, have a mass ratio of about 0.12 \citep{Brozovic2015}. Thus, the barycentre of the system lies between the two objects, and the regime of binary dynamics is most applicable.  Four circumbinary moons, Styx, Nix, Kerberos, and Hydra, have also been identified.  With the better characterisation of the \pc\ system stemming from the high-resolution view of \emph{New Horizons}, we can gain deeper insight into this system. This work aims to investigate two tracers of Pluto and Charon's formation: craters on the surface of Charon and debris that escaped into the Kuiper belt.

\cite{McKinnon1989}, \cite{Canup2005,Canup2011}, and others have proposed and refined a giant impact origin for the \pc\ binary.   \cite{Canup2011} studied a variety of collisions between Pluto and an impactor. The bodies can be either differentiated or non-differentiated; different incoming trajectories are simulated to understand the variations in the resultant system.  A giant collision of this type will typically form a moon, a disc, or both.  This study finds that a body one third to half the mass of the primordial Pluto will form Charon when it collides, although the newly formed moon tends to form with high eccentricity and  a pericentre close to Pluto (within a few Pluto radii).  If the impactor is differentiated, a disc is very likely and will have mass anywhere from 0.001\% of Pluto's mass to Charon's mass.  A post-collision disc may extend out to about 30 Pluto radii.  After the Charon forming impact, Charon is thought to migrate to its current position via tidal evolution. This tidal evolution can either be eccentric \citep{Cheng2014} or circular \citep{Dobrovolskis1989,Dobrovolskis1997,Peale1999} and should take at most a few million years. Charon concludes its migration tidally locked to Pluto with a 6.4 day period (semi-major axis of roughly 17 Pluto radii) and has eccentricity $\le 5\times10^{-5}$.

\subsection{Pluto's Moons}

Despite a compelling explanation for the formation of Charon, a theory for the emplacement of the four small circumbinary moons remains elusive.  Many works, such as \cite{Ward2006}, \cite{Lithwick2008a,Lithwick2008}, \cite{Canup2011}, \cite{Cheng2014a}, \cite{Kenyon2014}, and \cite{Walsh2015}, have tried to explain the location of the small moons.  Dynamical stability studies by \cite{Youdin2012} predicted low masses and high albedos for the moons, which were confirmed by \cite{Brozovic2015}, and \emph{New Horizons} \citep{Stern2015}. They find that the moons have masses of about $<1\times10^{-6}$, $3.1\times10^{-6}$, $1.1\times10^{-6}$, and $3.3\times10^{-6}$ relative to Pluto for Styx, Nix, Kerberos, and Hydra, respectively. These limits suggest that the circumbinary moons are icy, consistent with an origin in the disc from the Charon-forming impact. Nevertheless, many features of these moons remain difficult to explain when accounting for the tidal history of Charon. Specifically, the migration of Charon would easily destroy the extreme coplanarity ($<0.5$\deg), low eccentricity ($<0.006$), and the nearness to resonance (nearly 3:1, 4:1, 5:1, and 6:1 with Charon)  \citep{Brozovic2015}.   

The dynamical properties of the moons listed above are most consistent with in-situ disc formation, yet the discs in the \cite{Canup2011} simulations simply do not have enough material at the moons' current locations to form them.  Many proposed solutions have invoked resonant transport from the inner disc (where bodies form) to the outer disc, but these methods often pump the eccentricities and/or inclinations of the small moons well above the observed values.  The corotation resonance from \cite{Ward2006} would not excite eccentricities, but this method requires different Charon eccentricities to transport each moon.  Thus, \cite{Lithwick2008a} and \cite{Cheng2014a} suggest that this mechanism is unlikely. \cite{Cheng2014a} show a method to capture and transport disc material outward in a low (albeit non-zero) eccentricity orbit though capture into multiple Lindblad resonances while Charon is tidally evolving; however, they are unable to migrate material at the 3:1 and 4:1 commensurability with Charon (the locations of Styx and Nix).  \cite{PiresdosSantos2012} suggests that the current moons could come from collisions of other bodies near Pluto in the Kuiper belt, but the collision time-scales for massive enough objects are too long.  \cite{Walsh2015} suggest that the moons could form from disruption of an existing satellite in the system. This would provide a secondary disc, possibly at larger orbital radii, from which the moons can form, but still struggles to account for the wide range of circumbinary moon semi-major axes. 

\subsection{The Kuiper Belt and Collisional Families}

The history of the \pc\ system is tied to the history of the Kuiper belt and Kuiper belt objects (KBOs). A plethora of works beginning from \cite{Malhotra1993,Malhotra1995} have explored the early history of the \sols\ and the sculpting of the Kuiper belt via giant planet migration. In these scenarios, Neptune and Pluto begin closer to the Sun than they are today.  Neptune then migrates outward to its current orbit and Pluto is captured into the 3:2 resonance.  During this process, Pluto's orbit gains both eccentricity and inclination.

It is likely that the Charon-forming collision occurred early in the history of the \sols\ because the density of planetesimals was higher and thus collisional time-scales shorter. Additionally, works such as \cite{Levison2008} propose that there may be large numbers of larger objects (Pluto-sized) in the primordial Kuiper belt, which means that the cross section for giant impacts was larger.  Therefore, Pluto and Charon have likely existed in their current state for most of the Kuiper belt's history and should record information about the surrounding population of KBOs through cratering.  \cite{Greenstreet2015} simulates the expected crater size distribution on the surfaces of Pluto and Charon for both ``divot'' \citep[discontinuous double power law, e.g.][]{Shankman2016} and ``knee''  \citep[broken power law, e.g.][]{Bernstein2004,Fraser2014} Kuiper belt populations.  The true size distribution is still uncertain due to small samples and the likely presence of multiple populations.

Another interesting feature of massive KBOs is the presence of collisional families. Many KBOs, including the \pc\ system, show evidence of giant impacts that would produce a collisional family. The Haumea collisional family originally reported by \cite{Brown2007} is the only identified collisional family in the Kuiper belt.  This family consists of roughly a dozen objects with similar compositions and orbits to the dwarf planet Haumea.  In Haumea's case, the collisional family was easily identified because the members share a striking spectral feature and because the velocity dispersion of family members is about an order of magnitude lower than expected \citep{Schlichting2009}.  The typical collisional family, however, should have velocity dispersions closer to the escape velocity from the parent system, which is closer to one \kms.  \cite{Marcus2011} find that collisional families in the Kuiper belt are difficult to distinguish using the same method of low velocity dispersion used to find the Haumea family, but these families may be possible to pick out using other methods, such as clustering in inclination.  They also estimate that there should be, at most, a handful of collisional families from massive collisions and a few tens of families from progenitors of 150 km in size. The Haumea collisional family is suggested to be old (from less than 1Gyr after \sols\ formation) and therefore may be primordial \citep{Ragozzine2007}.  Thus, the majority of collisional families might stem from a time when the Kuiper belt was more dense, before dynamical stirring by Neptune. \cite{Leinhardt2010} note that collisional families in the Kuiper belt and the main asteroid belt have different characteristics due to Kuiper belt giant collisions tending to be slower and more massive.

\medskip
In this work, we investigate the evolution of a debris disc resulting from the Charon-forming collision. We look at collisions onto Charon's surface that might leave visible craters.  This crater population may contaminate measurements of the KBO size distribution.  We also look at the population of debris ejected into the \sols\ that might manifest as a Pluto collisional family in the Kuiper Belt.  In Section~\ref{CBD}, we discuss the circumbinary dynamics in the \pc\ system.  Section~\ref{method} presents our simulation methodology.  Section~\ref{charon} presents results for collisions onto Charon's surface, while Section~\ref{sols} shows the properties of ejected particles. 

\section{Circumbinary Dynamics in the Pluto System} \label{CBD}

The origin of Pluto's four circumbinary moons still remains a mystery.   The destabilizing influence of the binary almost certainly rules out in-situ formation if Charon undergoes eccentric tidal evolution.   As noted by \cite{Kenyon2014},\cite{Walsh2015}, and \cite{Bromley2015}, the \cite{Holman1999} binary instability boundary provides strict limitations on the stable locations of particles around the \pc\ binary.  The location of this empirical boundary, $a_{\textrm{crit}}$, is a function of both binary eccentricity and mass ratio ($\mu=M_{\textrm{C}}/(M_{\textrm{P}}+M_{\textrm{C}})$), as shown in equation~\ref{eq:stab}.  The overwhelming majority of particles that cross inside this boundary will go unstable in less than $10^4$ orbital periods and either eject from the system or collide with another body.
\begin{equation}\label{eq:stab}
\begin{split}
a_{\textrm{crit}}/a_{\textrm{PC}}=&1.60+5.10e-2.22e^2+4.12\mu-4.27e\mu\\
&-5.09\mu^2+4.61e^2\mu^2\\
\end{split}
\end{equation}

\begin{figure}
\centering
\includegraphics[scale=.4]{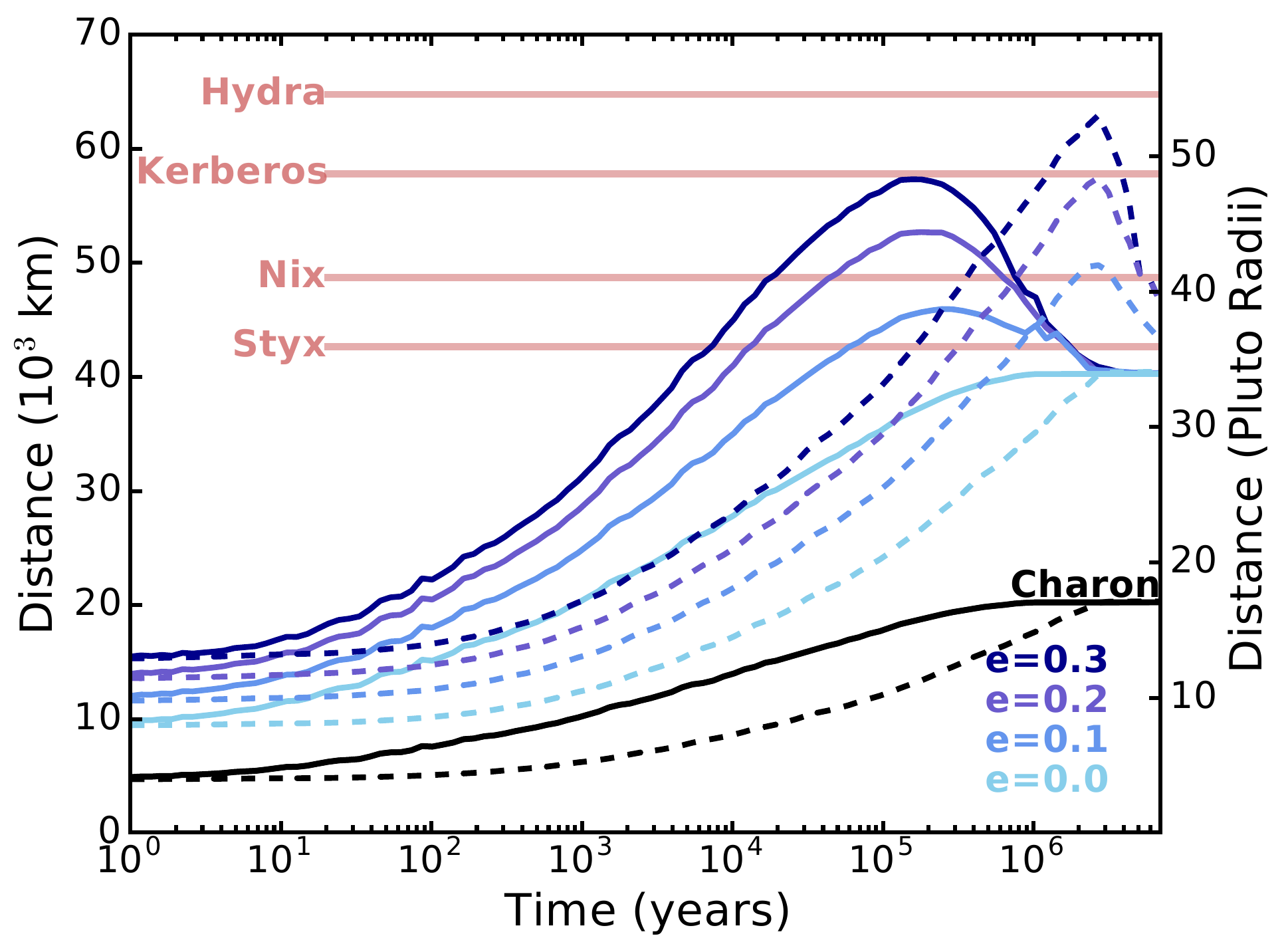}
\caption{ The location of the binary instability boundary as a function of time for different tidal evolution models.  The solid lines show the constant $\Delta t$ model and the dashed lines show the constant $Q$ model from \protect\cite{Cheng2014}.  The colour denotes initial eccentricity.  The black line shows the  semi-major axis evolution of Charon, and the red lines show the current locations of the four circumbinary moons.  In every case but the $e=0$ model, the instability boundary sweeps over a present-day location of one or more of the moons.  Bodies that cross the instability boundary will be driven unstable on short time-scales (less than 1000 years), so the moons cannot have been formed in situ if Charon  undergoes the tidal evolution presented here. 
\label{stab}}
\end{figure}

Simulations of Charon's formation suggest that it may have formed with an initial eccentricity as high as $e=0.8$.  Subsequently, Charon must undergo outward tidal evolution to reach its current semi-major axis and low eccentricity.  Tidal evolution models such as \cite{Cheng2014} require that Charon remain eccentric for nearly the entire outward migration.  If we apply the binary instability boundary in equation~\ref{eq:stab} to the Cheng models (both the semi-major axis and eccentricity evolution), we find that one or more of the circumbinary moons would be unstable for any tidal evolution model except one with zero eccentricity.  This is shown in Figure~\ref{stab}, where we plot the location of the instability boundary against time for different tidal evolution models.  The coloured lines show different initial eccentricities for the constant $\Delta t$ model (solid) and constant $Q$ model (dashed) (see \citealt{Cheng2014}  Sections 2.1 and 2.2 for a description of the  $\Delta t$ model and constant $Q$ model, respectively).  The semi-major axis evolution of Charon is shown in black, and the present-day locations of the four moons are shown in red. Debris or moons interior to (below) any of these curves cannot survive the tidal evolution of Charon because the instability time-scale is much shorter than the migration time-scale.  For instance, at the location of Hydra, the period is 38 days; $10^4$ orbital periods is just over 1000 years and is much shorter than the ~Myr migration time-scale for Charon.  Thus in situ moon formation from the initial debris disc is inconsistent with these eccentric tidal evolution models. It is possible to form the moons in situ if Charon's orbit is initially circular or if the eccentricity is damped early in the tidal evolution history.

Particles in the debris disc will encounter the instability boundary as it sweeps outward with the migration of Charon. The debris that interacts with the instability boundary will likely be ejected from the system or collide with Charon.  \cite{Smullen2016a} find that circumbinary planets initially exterior to the instability boundary will preferentially be ejected from the system when scattered toward the central binary by other bodies.   Compared to systems with a single central object, collisions are much more rare (by up to an order of magnitude) in a binary system.  Those objects that collide will more often collide with the less massive body as shown by \cite{Sutherland2016}.  This behaviour can be understood from simple three-body dynamical arguments discussed in Section 5.2 of \cite{Smullen2016a}. Thus, as Charon migrates outward, the instability boundary sweeps across the previously unperturbed disc and causes new waves of particle loss.

\section{Methods} \label{method}

We investigate the fates of debris in a disc from the formation of Charon.  First, we simulate the interaction of the \pc\ system with a disc of test particles in isolation and track the final outcomes of particles. In these simulations, we examine the impact of collisions onto the surface of Charon. We also record all ejected particles at Pluto's Hill sphere, and then we inject these ejected particles into the \sols\ and integrate to understand the long-term behaviour of this population of Pluto ejecta.  

We utilise integrators in the \cite{Chambers1997} \nb\ integrator \merc.  For integrations of the \pc\ system itself, which has two massive bodies, we use the Gauss-Radau variable time step integrator \emph{Radau} in the modified version of \merc\ presented in \cite{Smullen2016a}. We use the Bulirsch-Stoer integrator \emph{BS} for the forced migration simulations of the \pc\ system. For integrations of particles in the \sols, we use the \emph{Hybrid} integrator from the unmodified version of \merc, which uses a Bulirsch-Stoer integrator for close encounters and a symplectic integrator for all other time steps.

Both the isolated \pc\ and the \sols\ integrations have been effectively parallelized by running smaller sets of test particles with the same sets of massive bodies.  This means, though, that our simulations do not finish with identical ending conditions for the massive bodies because we allow close encounters with test particles.  When \merc\ uses an adaptive time step method, such as the Bulirsch-Stoer or \emph{Radau} integrators, the time step of the simulation changes.  In the \sols\ integrations, the overall time of the simulation returns to the symplectic time step after the close encounter has passed, but an imprint of the change remains. Minute variations in the orbits during adaptive stepping routines act as a source of ``chaos'' in the system, leading to significant variations in the final conditions after long time scales.  Because of the many massive bodies and long time-scales in the \sols\, these variations manifest themselves in all orbital elements.  In the \pc\ system, where there are only two massive bodies, the variations tend to manifest as differences in final mean anomaly.

\subsection{\pc\ System}

To reach its current position, the \pc\ binary must have migrated, but the details of this migration are unknown.  To this end, we implement three unevolving \pc\ orbits representing different stages of eccentric tidal evolution, and we also implement a circular migration model.  

For the constant orbit integrations, our initial conditions are motivated by the \cite{Cheng2014} tidal evolution models for Charon: we initialise Charon at binary semi-major axis $a=5$\rp\ and eccentricity $e=0.3$, $a=17$\rp\ and $e=0.3$, and $a=17$\rp\ and $e=0.0$.  Thus, we cover a set of \pc\ orbits that span the most compact to the widest at a plausible range of eccentricities. 

Each integration begins with Charon at a mean anomaly $M$ of 90\deg\ and is integrated to $10^7$ days, or roughly 1.5 million \pc\ orbits at $a=17$\rp. The disc has 27060 test particles that range in barycentric distance from 0--65000 km with eccentricities randomly drawn from a uniform distribution from 0 to 0.01 and inclinations randomly drawn from a uniform distribution from 0 to 0.5\deg.  \cite{Walsh2015} find that a stable ring around the \pc\ binary should collisionally circularize in about a decade, meaning that any free eccentricity in the debris' orbit set from the progenitor collision should damp within a few decades.  While the disc will not be perfectly circular due to a forced eccentricity from the binary, in the regions under consideration, the forced eccentricity will be less generally much less than 0.3 using the formulation from \cite{Leung2013}.  \cite{Mudryk2006} note that the instability of circumbinary material is not a strong function of initial eccentricity, so an initial lack of forced eccentricity will not impact our results.   The disc is then evolved under the influence of the binary, so within very few orbital periods, the inner edge of the disc (just exterior to the binary instability boundary) becomes slightly eccentric due to the forced eccentricity from the central binary.  The test particles are drawn such that the average spacing between any two particles is constant; this is achieved by initialising the same number of test particles in 132 annular rings where the area of each ring is constant.  Consequently, the innermost rings have widths of several hundred to a few thousand km, while outer rings will only be a few km in width.  Our disc is unphysically large for the \cite{Canup2011} models of \pc\ formation, but such a large simulated disc allows us to convolve any physical disc model in post-processing.  While our disc is initialised around the \pc\ barycentre, we tested a model with a Pluto-centred disc and found little change in particle fates.

We also implement the migration model described below in Section~\ref{method:mig} in the \pc\ system to understand the differences in particle fates caused by an evolving orbit. Charon migrates through the disc (which is the same as above) from 5\rp\ to 17\rp.  We run the simulations for $10^5$ yr (about three times longer than the constant orbit simulations). 

We set the sizes of Pluto and Charon using spherical shapes and densities of 1.88 \gcm\ and 1.65 \gcm, respectively.  Particles are tracked to the surface of the massive bodies by setting the massive body close encounter radius to one physical radius, thereby ensuring that there are no extrapolation errors introduced into collision rates.  Particles are considered to be ejected when they reach a distance equivalent to Pluto's modern-day Hill sphere of about 0.06 AU (about 140 times the initial disc extent in the simulations).  We track Pluto-barycentric positions and velocities at the ejected time step to use in our \sols\ integrations.

As a test, we also integrated the present \pc\ system with the four circumbinary moons and the test particle disc.  The presence of extra massive bodies in the system results in extra particle loss, as the circumbinary moons help scatter debris inwards.  Most of these losses are through ejections instead of collisions.  Additionally, significant structure appears in the disc, such as co-orbital debris near the small moons and shepherded rings between the moons.  We choose to not analyse these simulations in detail because we are concerned with the impact of the \pc\ binary alone. Additionally, due to the uncertain nature of the origin of the circumbinary moons, there is no way to estimate the appearance of the system at our different \pc\ configurations.

At $10^7$ days, our constant orbit simulations show median energy conservation of $\Delta E/E=1\times10^{-11}$ and median angular momentum conservation of $\Delta L/L=1\times10^{-11}$. 

 We integrate the \pc\ system in isolation despite the potentially significant perturbations from the Sun on long time-scales.  There are two major effects the Sun could have on debris in the outer parts of Pluto's Hill sphere: an induced harmonic oscillation in specific angular momentum due to solar torques and secular perturbations causing Kozai oscillations.  Following the example of \cite{Benner1995} and \cite{Goldreich1989}, the Sun should drive a change in the specific angular momenta of disc particles with a period equal to half of Pluto's heliocentric orbital period, or about 124 years. The change in the mean specific angular momentum simplifies to
\begin{equation}\label{eq:delh}
\frac{\delta h}{\bar{h}}=\frac{15}{8}\frac{P_{\textrm{\tiny particle}}}{P_{\textrm{\tiny Pluto}}} \frac{e^2_{\textrm{\tiny particle}}}{\sqrt{1-e^2_{\textrm{\tiny particle}}}}
\end{equation}
where $P_{\textrm{\tiny particle}}$ denotes the period of the particle in the disc, $P_{\textrm{\tiny Pluto}}$ is the heliocentric period of Pluto, and $e$ is the eccentricity of the disc particle.  For a disc particle with $e=0.9$ at 1000 Pluto radii from the barycenter (an orbital period of $\sim$8 years), this constitutes only a 10\% change in the angular momentum every century; for a disc particle with the same eccentricity at 100 Pluto radii, the difference is less than 0.5\%.  Our simulations do not produce a population of high apocentre bodies that remain in the simulation for more than about 200 years because things are scattered out of the system very quickly.  Thus, this induced oscillation from solar torques is unlikely to impact our results.  Similarly, Kozai oscillations induced by the Sun will not affect the outcomes of our simulations.  The Kozai timescale for \pc\ and a test particle as the inner binary and the Sun as the outer component,  is a few to a few thousand times Pluto's heliocentric orbital period, depending on the period of the test particle \citep{Fabrycky2007}.  For the average disc particle, the Kozai timescale will be longer than the length of our simulations.  While both of these effects will change the orbits of any debris that remains in the system over long time-spans, the fates of particles quantified in this work should not be strongly influenced by the Sun prior to leaving the Pluto-Charon Hill Sphere.  

\subsection{Solar System} \label{method:sols}

We inject particles ejected from \pc\ into the \sols\ at three different points in the \pc\ heliocentric orbit: $M_{\textrm{P}}=180$\deg\ (apocentre),  $M_{\textrm{P}}=90$\deg, and  $M_{\textrm{P}}=0$\deg\ (pericentre).  We also simulate test particles in an evolving \sols\ using the migration model described below in Section~\ref{method:mig}.  We use a \sols\ model comprised of the Sun (with mass increased by the masses of the four terrestrial planets), the four giant planets, and Pluto. The positions and velocities of the planets and Pluto are taken from the JPL {\sc horizons} catalogue.\footnote{Apocentre in the {\sc horizons} catalogue occurs on 2114 Feb 22, $M=90$\deg\ is on 2051 Dec 5, and pericentre is on  1990 Jan 29.}  We randomly sample 16000 test particles, distributed over 100 individual simulations, from the full set of particles ejected from the simulation in which Charon has $a=17$\rp\ and $e=0.3$.  The test particles are boosted into the \sols\ frame from the isolated \pc\ system by adding the \pc\ barycentre position and velocity at the start of the simulation.  We run two orientations of the \pc\ disc with respect to the \pc\ heliocentric orbit.  The first set is aligned with Pluto's heliocentric orbit, while the second set is aligned with the present-day Pluto-Charon orbit, having $i=96.3$\deg, $\Omega=223.0$\deg, and $\omega=172.6$\deg\ with respect to Pluto's heliocentric orbit.\footnote{These angles are taken from the JPL {\sc horizons} catalogue.}  We set the time step of the simulations to be 200 days, and the hybrid changeover is set to 3 Hill radii.  Planetary radii are calculated using spheres with densities 1.33, 0.70, 1.30, 1.76, and 1.88 \gcm\ for Jupiter, Saturn, Uranus, Neptune, and Pluto, respectively.  The ejection radius is set to 2000 AU.  We integrate the simulations for 1.5 Gyr, or about 6 million heliocentric Pluto orbits.

At 1.5 Gyr, our median energy conservation is $\Delta E/E=1\times10^{-6}$ and our median angular momentum conservation is $\Delta L/L=4\times10^{-12}$.

\subsection{Migration} \label{method:mig}

 We implement the \cite{Malhotra1995} migration model in both the isolated \pc\ system and the \sols.  In the model, migration is considered a drag force.  This drag acceleration takes the form
\begin{equation}\label{eq:mig}
\pmb{a}_{\textrm{migration}}=-\frac{\pmb{\hat{\textrm{v}}}}{\tau} \left[ \sqrt{\frac{GM_\odot}{a_{\textrm{final}}}} -\sqrt{\frac{GM_\odot}{a_{\textrm{init}}}} \right] \exp\left(-\frac{t}{\tau} \right).
\end{equation}

In the \sols\ integrations, the acceleration is applied to each of the giant planets (Pluto is allowed to migrate naturally under the influence of Neptune).  We choose $\tau=2\times10^6$ years our migration time-scale, and we use the initial positions, eccentricities, and migration distances ($\Delta a \equiv a_{\textrm{final}}-a_{\textrm{init}}$) given in Table~\ref{tab:mig}. The orbital angles for the giant planets are the same as in the pericentre simulations, and inclinations are set to the modern-day values.  Pluto's inclination is set to 0\deg\ relative to the plane of the \sols.  These values are all consistent with those used in  \cite{Malhotra1995}.  While this migration model is not as involved as those in more modern works  \citep[e.g.,][]{Levison2008}, it is adequate to explore some of the impact of a dynamically evolving \sols\ on the orbits of Pluto ejecta.  We use only this migration model as it is simple to implement; a more complicated migration model is beyond the scope of this paper.  

\begin{table}
\centering
\caption{ Initial conditions for migration: We show the initial semi-major axis, prescribed change in semi-major axis, and initial eccentricity for the massive bodies in our simulations.  Pluto has no $\Delta a$ because it is allowed to naturally be captured into the 3:2 resonance. } 
\label{tab:mig}
\begin{tabular}{lcrc}
\hline
Planet & $a_{\textrm{init}}$  & $\Delta a$ & $e_{\textrm{init}}$   \\
 &  (AU) & (AU) &    \\
\hline
Jupiter 	& 5.4  & $-0.2$ & 0.048 \\
Saturn	        & 8.7  &  0.9 & 0.056 \\
Uranus  	& 16.3 &  2.9 & 0.046 \\
Neptune 	& 23.2 &  6.9 & 0.009 \\
Pluto 	        & 32.0 &  --- & 0.050 \\
\hline
\end{tabular}
\end{table}

\begin{figure}
\centering
\includegraphics[scale=.45]{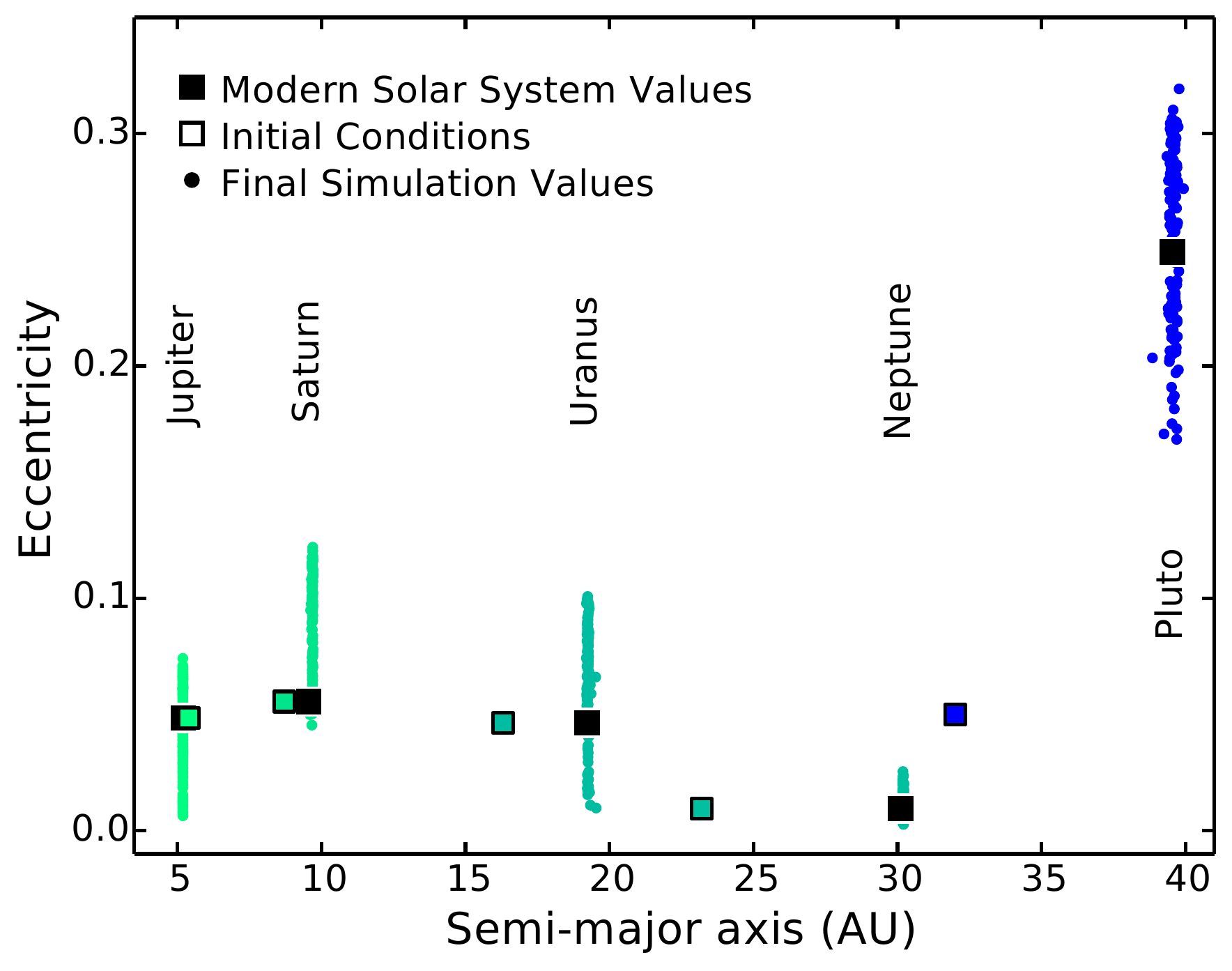}
\caption{ The  combined final eccentricities vs. semi-major axis for the giant planets and Pluto after 1.5 Gyr in the two sets of simulations that include migration.  The colour denotes the body, and the large black squares show the modern-day values for each body.  The coloured square outlined in black shows the initial conditions in our simulations and can also be found in Table~\ref{tab:mig}.  Pluto migrates under Neptune's influence to its proper semi-major axis and eccentricity.  The mean semi-major axis for all simulated bodies is less than 1\% different from the actual values.  The eccentricities have larger scatter but are consistent with other \sols\ migration models, as are the inclinations. 
\label{mig}}
\end{figure}

We use the same hybrid integrator as we did for our non-migrating \sols\ simulations, and we adopt the same orbital angles for the giant planets and Pluto in the migration simulations as we do in the pericentre run above in Section~\ref{method:sols}.  We again use 16000 initial test particles distributed over 100 simulations and run for 1.5 Gyr.  The first 100 Myr of each simulation are run with a time step of 100 days, and the final 1.4 Gyr are run with a time step of 200 days.  Figure~\ref{mig} shows the final semi-major axes and eccentricities of the massive bodies (the four giant planets and Pluto) in the simulations at the end of 1.5 Gyr.  The expected values are shown by the large black squares, and the initial values are shown by the coloured squared outlined in black.  60 of 100 runs  with the disc aligned with Pluto's orbit and 66 of 100 runs with the disc aligned with the modern \pc\ orientation finished successfully with Pluto in the 3:2 resonance; these are the only ones presented in Figure~\ref{mig} and considered for the rest of the analysis.  Most of the unsuccessful runs saw Pluto ejected from the \sols.  All semi-major axes for the massive bodies are within 1\% of the expected semi-major axes.  The eccentricities vary much more, but do match the ranges encompassed by other \sols\ migration models, such as those presented in \cite{Tsiganis2005}.  The final inclinations are also very close to modern-day values.

In the isolated \pc\ system, the only body to which the migration drag force is applied is Charon. The migration time-scale is set to  $\tau=10^4$ yr.  We use the Bulirsch-Stoer integrator in the modified \merc\ and integrate for $10^5$ yr.  A migration time-scale of $10^4$ years is short for proposed tidal evolution models for Charon, but, due to computational limitations, we choose a shorter time. However, the dynamics will scale similarly with longer tidal evolution time-scales because the dynamical time-scale of test particles in these regions is very short compared to the speed with which Charon moves radially outward. Particles inside the binary instability boundary will be removed within about $10^4$ orbital periods; at the instability boundary of the initial \pc\ orbit, this is around 100 yr, which is much shorter than the time it takes for Charon to migrate though the region.

\section{Fate of Debris: Collisions with Charon} \label{charon}

\subsection{Relevant Time-scales}\label{coll:time}

The tidal evolution of Charon should take on the order of 1 Myr \citep{Dobrovolskis1989,Cheng2014}.  Collisions can occur at any time during or shortly after the migration as the binary instability boundary excites disc material. Observations from \emph{New Horizons} presented in \cite{Singer2016} suggest that Charon's surface age is upwards of 4 Gyr and could stem from shortly after formation, so these collisions should be preserved.

Because Charon is formed via a violent collision, the surface should not be solid early in the binary's history.  Collisions would therefore not be preserved in this era.  Thus, we must estimate a time at which collisions would be recorded. If we turn to the surface cooling time-scales of non-tidally heated bodies in the \sols, such as an estimated $10^3$ years for an atmosphereless Mars from \cite{Monteux2015} or $10^4$--$10^6$ years for Earth from \cite{Spohn1991}, \cite{Tonks1993}, and others, we can make a rough estimate that the much smaller and icier Charon cooled on time-scales of a few hundred to a few thousand years in the absence of other effects.  The surface cooling time-scale is therefore much smaller than the tidal evolution time-scale.  Accordingly, a significant fraction of the disc may be unperturbed when Charon solidifies and the continued accretion of disc material should be imprinted on the surface.  Note that this cooling time-scale does not account for sources of internal heating, which we discuss in Section~\ref{coll:tide}.

\subsection{Collisions and the Disc} \label{coll:disc}

\begin{figure}
\centering
\includegraphics[scale=.45]{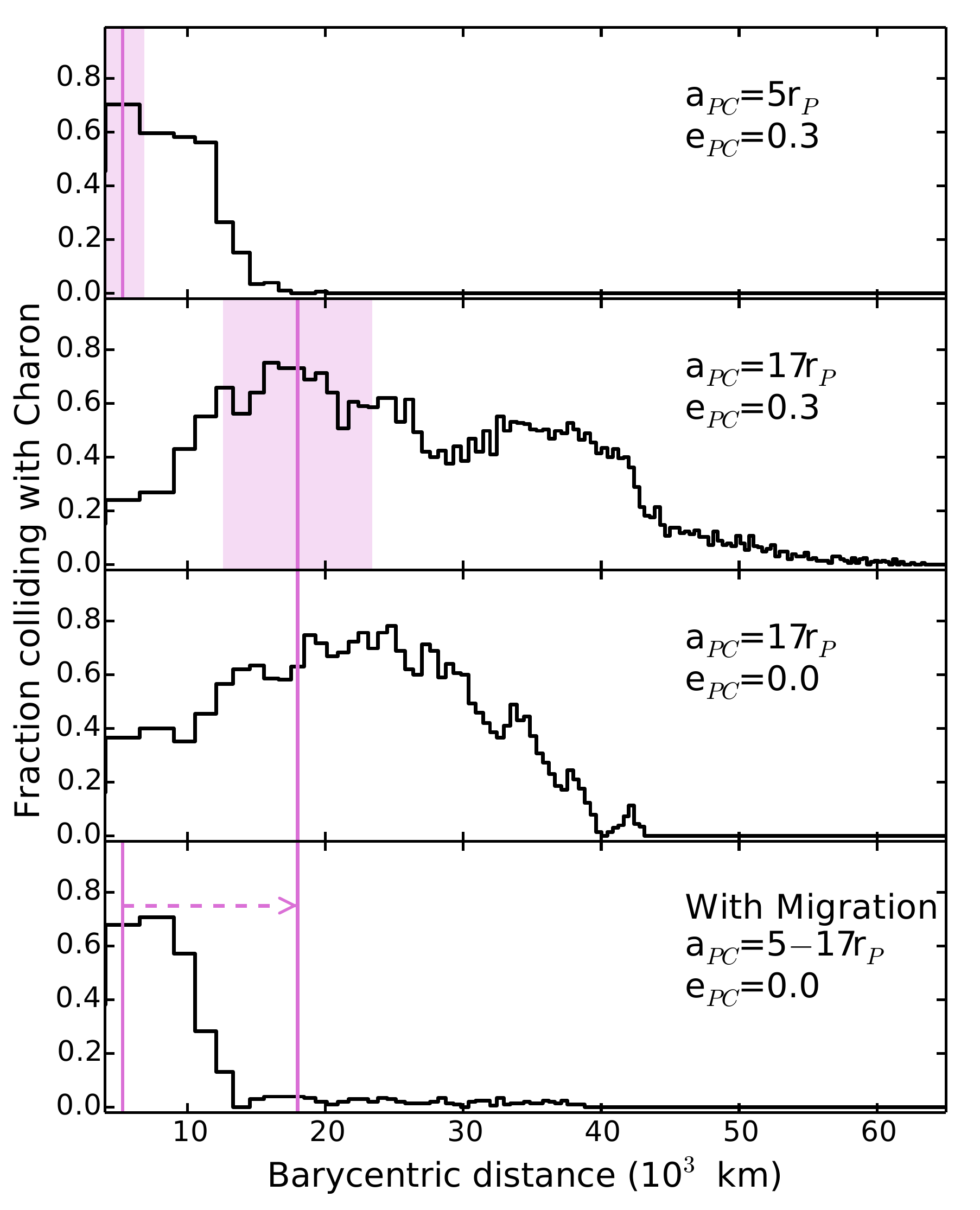}
\caption{ The fraction of disc particles colliding with Charon per radial bin as a function of barycentric distance.  The four panels show different \pc\ orbits: $a=5$\rp\ and $e=0.3$, $a=17$\rp\ and $e=0.3$,  $a=17$\rp\ and $e=0$, and a migration from 5--17\rp\ with zero eccentricity.  The purple line shows the semi-major axis of Charon, and the purple rectangle shows the radial extent of Charon's orbit (pericentre to apocentre).  In some regions, upwards of 50\% of material collides with Charon. The outer edge of collisions is governed by the instability boundary; exterior to this boundary, material is unperturbed.
\label{colls}}
\end{figure}

Figure~\ref{colls} shows the colliding fraction as a function of barycentric distance for four \pc\ orbits: $a_{\textrm{PC}}=5$\rp\ and $e=0.3$, $a_{\textrm{PC}}=17$\rp\ and $e=0.3$,  $a_{\textrm{PC}}=17$\rp\ and $e=0$, and a migrating Charon from $a_{\textrm{PC}}=5$--17\rp\ on a circular orbit.  The first three are representative of three phases of a proposed tidal evolution model for Charon from \cite{Cheng2014}, while the last actually moves Charon through the disc.  The most compact configuration is similar to the orbit at $\sim$100 years after impact. The wide, eccentric configuration is most similar to what is expected around $10^4$--$10^5$ years after impact and is where we see the most disc disruption take place.  Finally, the wide, circular case is similar to the \pc\ system seen today.   The semi-regular decreases in colliding fractions in the circular case (third panel) arise where low-order mean motion resonances (such as the 3:2, 2:1, 5:2, and 3:1) cause preferential ejection of material. This circular case is least realistic, as the wide, eccentric Charon would have previously cleared out all of the debris able to interact with a wide, circular Charon.

While the bulk of collisions should occur early, there will be craters from disc debris throughout the migration. Both the most compact case (top) and the migrating case (bottom) show a sharp decrease in collisions outwards of about 13000 km; the collisions interior occur on time-scales of a few to a few hundred years, which is comparable to the solidification time of Charon. Thus, neither of the collision populations interior to 13000 km would be visible on the surface of Charon.  However, in both cases, a significant quantity of disc material remains exterior and interacts with the evolving binary at a later time.

\subsection{Craters on Charon's Surface}

We can estimate the number of craters visible on the surface of Charon for a given disc profile and size distribution of disc particles.  We calculate cratering for two \pc\ configurations: the $a=5$\rp\ and $e=0.3$ disc, as this is the stage with the most dynamical evolution of the disc, and the migration model.  Additionally, we only take the material exterior to Charon's orbit (outside the purple region in Figure~\ref{colls}) for the constant orbit model because the binary instability boundary would reach this location at roughly 1000 years in the \cite{Cheng2014}  constant $\Delta t$ model (see Figure~\ref{stab}); 1000 years is approximately the estimate for Charon's surface cooling time-scale. We only consider collisions originating outside of 14000 km in the migration model because the instability boundary will take about 1000 years to move to this location. Typical collision speeds should be roughly the escape velocity from Charon added in quadrature with the relative velocity of the collider and Charon, or about 0.5--1\kms.  This velocity of about 0.5 \kms\ is  about a quarter of the expected velocities of 1--2 \kms\ for KBOs impacting Pluto and Charon quoted in  \cite{Greenstreet2015}.  We use an impactor-to-crater size ratio of 5, which is a small but consistent value calculated using Charon's escape velocity in Greenstreet's equation 5.  \emph{New Horizons} has a resolution of 1--1.5 km on the surface of Charon for the largest data sets \citep{Moore2016}, so we take ``observable'' craters to be those larger than  3 km, implying an impactor at least  600 m in diameter.

\cite{Canup2011} finds that a debris disc from a Charon-forming impact can extend up to 30 Pluto radii and will range in mass from $10^{20}$--$10^{24}$ g.  We adopt a disc with radial extent of 30 Pluto radii and mass $10^{22}$ g for our analysis, an optimistic but not unrealistic disc. We convolve this disc model with the constant density, large radius test particle disc in our \nb\ simulations.  We then can calculate a disc surface density parametrized by the disc surface density index $\beta$ and radial extent $r$  over a Pluto radius $r_{\textrm{P}}$ such that
\begin{equation}\label{eq:sd}
\Sigma(r)=\Sigma_0\left( \frac{r}{r_{\textrm{P}}} \right)^{-\beta}.
\end{equation}
$\Sigma_0$ is a normalisation factor calculated by equating the disc mass and the integrated surface density.   By multiplying the surface density profile from the giant impact-motivated disc with our collision fractions from our simulated disc, we can calculate a colliding mass per radial bin.  Then, we must assume a particle size distribution to find a number of colliders.  We take this to be a power law parametrized by particle size index $q$.  Thus
\begin{equation}\label{eq:size}
N(s)=N_0\left( 2s \right)^{-q},
\end{equation}
where $s$ is the particle radius and $N_0$ is found by assuming that each body is a sphere with density $\rho=1$\gcm, an icy composition, and that our particles range in size from 1 cm to 5 km.  We choose the upper limit of $s=5$ km so that our debris is  equal in size or smaller than the existing moons.  Decreasing either the upper or lower size limit increases the number of visible craters, while increasing either limit decreases visible craters because more of the fixed mass goes into larger bodies, dropping the total number of colliders.  Finally, we sum the number of colliders in all radial bins as a function of size and calculate the number with radii greater than 300 m.  These are the impacts from the disc that would be visible on the surface of Charon with \emph{New Horizons} imaging.

\begin{figure}
\centering
\includegraphics[scale=.45]{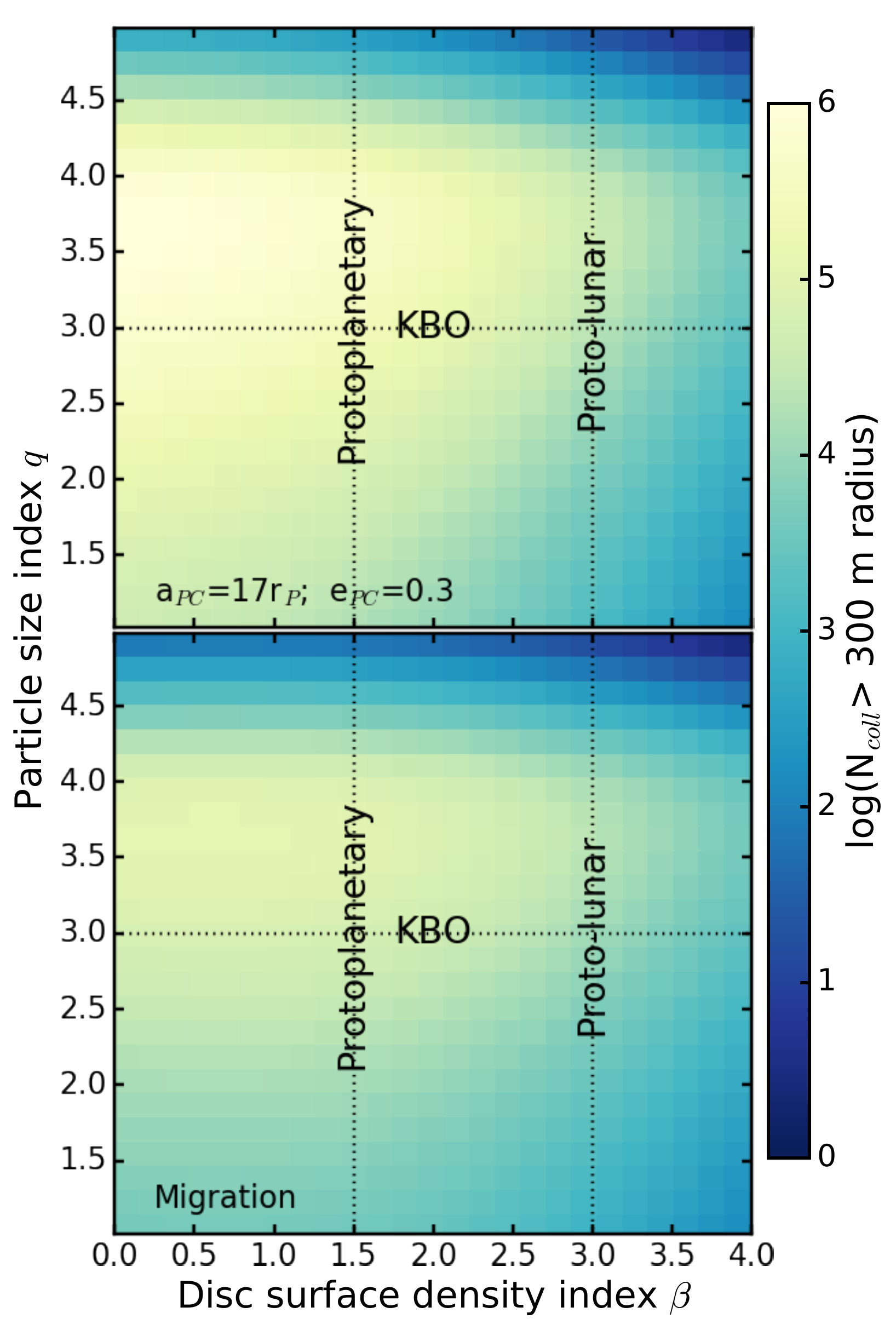}
\caption{ The number of  colliders  that would leave visible craters as a function of particle size index $q$ and disc surface density index $\beta$.  The parent disc extends to 30\rp\ and had a mass of $10^{22}$ g. The top panel shows the number of colliders for Charon with static orbit $a=17$\rp\ and $e=0.3$ and the bottom panel shows the same for the migrating Charon.  Only bodies external to the maximum extent of Charon's orbit are considered in the eccentric case and external to the instability boundary at 1000 years, located at 14000 km, in the migrating case (see the second and fourth panels of Figure~\ref{colls}, respectively).  Colliders are taken to be observable if they are greater than  600 m in diameter, which should correspond to a crater at least 3 km in diameter (3 km is twice the resolution of \emph{New Horizons}).  Also labelled are typical values of $\beta$ for protoplanetary and proto-lunar discs.  The $q$ value for Kuiper belt objects is taken from crater size measurements on the surface of Charon from \protect\cite{Singer2016}.  For reasonable values of $q$ and $\beta$, there should be  hundreds to a few tens of thousand craters on Charon's surface from the Charon-forming disc. 
\label{ncoll}}
\end{figure}

Our estimates for the number of colliders as function of the particle size index $q$ and disc surface density index $\beta$ are shown in Figure~\ref{ncoll} for two of the \pc\ orbits we simulate.  We take the $a=17$\rp\ and $e=0.3$ simulation as the most optimistic case (most collisions) and the migration model as the least optimistic (fewest collisions) for simulations in which Charon is near its current orbit.  We show typical values of $\beta$ for protoplanetary discs (which normally range from 0.5--1.5) and proto-lunar discs \citep[e.g.,][]{Charnoz2015}.  The value of $q=3$ labelled KBO is taken from crater size measurement on Charon from  \cite{Singer2016}.  At this value of $q$, we estimate that there may be  hundreds to hundreds of thousands of craters on the surface of Charon that stem from the Charon-forming disc. We expect the craters to be evenly distributed across Charon's surface because Charon was not initially tidally locked to Pluto and the height of the disc at the instability boundary (where the colliders typically originate) is comparable to the size of Charon. However, these craters  from the disc would be among the oldest on the surface and among the smallest with the range of allowed particle sizes we have chosen.   \cite{Singer2016} note that there appears to be a lack of small craters on Charon's surface, which suggests fewer impacts originating from the disc itself.  The real disc is therefore likely comprised of smaller debris (less than a few km in radius), smaller in radial extent, or has a steeper surface density index.   

If it is possible to identify and date craters on Charon as stemming from the disc, clues as to Charon's tidal evolution outward can be inferred.  Encounter velocities tend to be lower when Charon is in an eccentric orbit because the encounters are more likely to occur in the outer part of the orbit where orbital velocities are slower. Thus, if we can assume a similar size of impactors through all stages of the tidal evolution, epochs of an eccentric Charon would show smaller craters relative to a circular Charon.

Note that we do not consider collisions with Pluto for two important reasons: first, collisions with the more massive component of a binary are intrinsically rarer.  We see a factor of 3--4 reduction in the number of collisions with Pluto compared to collisions with Charon.  Secondly, the resurfacing time-scale of Pluto (specifically, Sputnik Planum) has been estimated to be less than 10 Myr by \cite{Moore2016} and \cite{Trilling2016}. The surface is therefore expected to be much younger than Charon, although some regions may be much older and thus susceptible to disc cratering. Any craters originating from disc debris in young areas would be erased by recent or previous resurfacings.

\subsection{Internal Heating and Charon's Surface} \label{coll:tide}

 In the previous sections, we did not account for the effects of internal heating, either through tidal or radiogenic processes. These two effects might  prolong the era of a liquid surface on Charon or cause slow resurfacing. 

\cite{Jackson2008} parametrize the tidal heating rate per surface area (their equation 3) as
\begin{equation}\label{eq:tid}
h=\left( \frac{63}{16\pi} \right) \frac{\left( GM_{\textrm{P}} \right)^{3/2}M_{\textrm{P}}R_{\textrm{C}}^3}{Q^\prime_{\textrm{C}}}a^{-15/2}e^2.
\end{equation}
Here, P denotes Pluto and C denotes Charon, which have been changed from Jackson's S (star) and P (planet), respectively. $Q^\prime$ is the tidal dissipation parameter and is given as $Q^\prime=3Q/\left(2k_2\right)$.  $Q$ is the tidal dissipation function (taken here to be 100), and $k_2$ is the Love number.  In the analysis, \cite{Jackson2008} consider anything with $h>2$ to be highly volcanic, $2>h>0.04$ to have enough heating for tectonic activity, and anything less than this to have too little internal heating to promote activity (what we refer to as dead).  For reference, Jupiter's moon Io has $h$$\sim$$2-3$ W/m$^2$, while Europa may have tidal heating as high as $h$$\sim$$0.2$ W/m$^2$.  Earth's heating, which comes from radiogenic sources and the heat of formation, is about 0.08 W/m$^2$.  We do not expect a surface in the volcanic regime to retain any craters as resurfacing is very fast.  Craters may be retained if the body is tectonic, although relaxation of the surface material  may make the craters smooth out or fill in over time. Craters on dead bodies should not undergo significant evolution without outside influence.

\begin{figure}
\centering
\includegraphics[scale=.45]{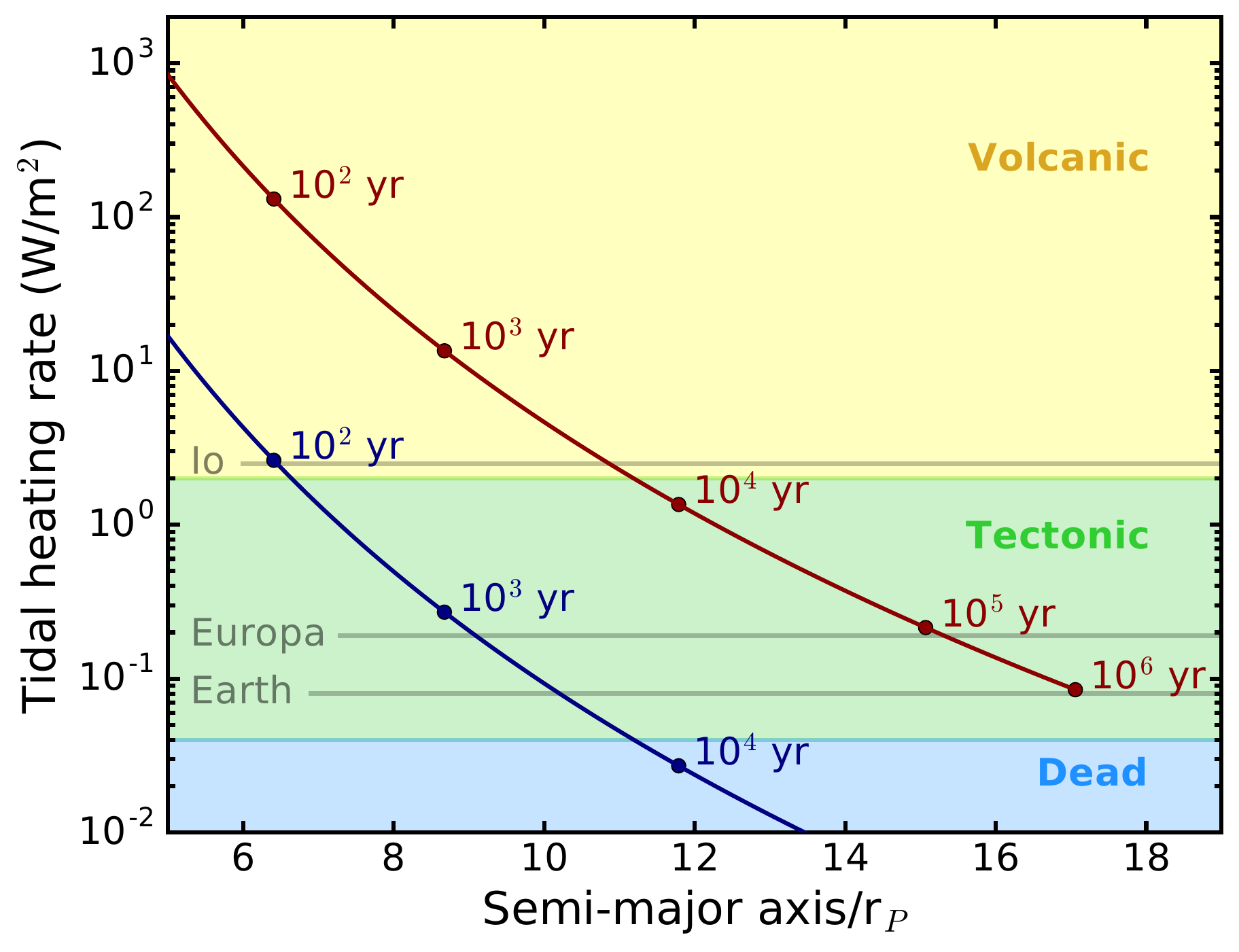}
\caption{ The tidal heating rate per unit surface area for Charon's tidal evolution history.  We apply the \protect\cite{Cheng2014}  constant $\Delta t$ semi-major axis evolution at constant $e=0.3$  to equation~\ref{eq:tid}.  The blue line shows the tidal heating rate for Charon with $k_2=0.006$ from \protect\cite{Murray1999} Table~4.1, and the dark red line shows the tidal heating rate for a Charon with an Earth-like Love number of $k_2=0.3$.  The yellow region shows what \protect\cite{Jackson2008} consider volcanic, or violently disturbed from tidal heating, green shows the tectonic regime where there may be some surface activity, and blue shows heating rates that lead to a geologically dead body.  Io, Europa, and Earth have been labelled for reference.  For the rocky Love number, Charon's surface would not cool enough to retain craters (cross the volcanic-tectonic boundary) until nearly $10^4$ years after formation.  As the binary instability boundary reaches the outer extent of the debris disc at around 5000 years after formation, we would not expect to see any craters from the debris disc.  With the icy Love number, Charon should not experience significant tidal heating after a few hundred years, so craters from the debris disc should be preserved.
\label{tidal}}
\end{figure}

Figure~\ref{tidal} presents $h$ vs $a$ for an eccentric Charon.  We adopt the  constant $\Delta t$ semi-major axis evolution from \cite{Cheng2014} (the  solid black line in Figure~\ref{stab}) and a constant eccentricity of 0.3. The two lines denote different values of the Love number $k_2$.  Blue shows an estimate of Charon's Love number from \cite{Murray1999} Table~4.1, who estimate $k_2=0.006$, and red shows a rocky, Earth-like Love number of 0.3.  We colour regions based on the \cite{Jackson2008} demarcations: yellow shows the volcanic region where tidal heating causes violent and fast resurfacing, green shows the tectonic region where a solid surface may be slowly changed over time, and blue shows geologically dead bodies. We also plot reference values for Io, Europa, and Earth.  The factor of 50 difference between the two Love numbers has a large impact on the expected observability of craters from the disc. With a large Love number, Charon spends the majority of the early evolution firmly in the volcanic region.  It is therefore unlikely that the surface would retain any craters as constant resurfacing is probable.  With the lower, and likely more appropriate, Love number, the time spent undergoing violent tidal heating is less than our estimated cooling time-scale in Section~\ref{coll:time}.  Thus, craters should be imprinted as soon as the surface cools from formation.

For any craters to be visible, the surface needs to be solid and slowly changing (for our purpose, in the tectonic regime) before the binary instability boundary reaches the outer edge of the disc.  Assuming the same disc extent of 30\rp used above and using the $e=0.3$ binary instability boundary in Figure~\ref{stab}, the surface of Charon needs to solidify by, generously, about 5000 years after formation to retain any craters from the debris disc.  We can change the time spent in the volcanic state most easily through $k_2$, as shown above, or through eccentricity. If Charon begins with a lower eccentricity, the time that Charon spends in the volcanic regime is much shorter because tidal heating scales as $e^2$.  For $e=0.1$ and $k_2= 0.3$, Charon moves into the tectonic regime by about 800 years after formation.  For a circular migration scenario, Charon undergoes no tidal heating.

In the tectonic regime, craters may last on the surface for thousands to millions of years.  \cite{Moore1998} fin that craters on the surface of Europa may last for up to $10^8$ years.  Because Charon should be absolutely cold by the end of the tidal evolution at $\sim$10$^6$ years, craters from the tectonic regime may still be visible.  They may show similar features to impact craters on Europa noted by \cite{Moore1998,Moore2001} such as shallow basins or relaxed crater walls.  In extreme cases, the craters may resemble more of a circular ridge than a true crater.  This may be a way to distinguish craters from the debris disc from KBO craters: as debris craters will be among the oldest and may stem from a time when Charon had different surface properties, the physical appearances of the two crater populations may be very different. 

The other important source of internal energy, radiogenic heating, is orders of magnitude smaller but may have an impact on Charon's surface a few Gyr after formation.  \cite{Desch2009} run a full radiogenic heating model for Charon and find that the heat flux through Charon's surface peaks at 5 mW/m$^2$ at 0.5-1.5 Gyr after formation.  This is enough heat flux to differentiate the interior of Charon but cannot melt the crust, which is expected to be 60--85 km thick \citep{Rubin2014}.  Thus, once the surface of Charon has solidified, further heating should not completely resurface the moon; this is consistent with surface age measurements from \cite{Singer2016}.  Radiogenic heating is enough to make the surface malleable and allow long-term relaxation of craters.  Additionally, radiogenic heating may drive cryovolcanism on Charon.  Cryovolcanism could contribute to minor resurfacing by filling in craters and causing erosion. \cite{Desch2009} estimate that around 120 m of ice will be deposited uniformly on the surface through cryovolcanism over the past 3.5 Gyr, although later studies such as \cite{Neveu2015} suggest that this may be an overestimate of the cryovolcanic activity because the surface may be difficult to crack.  While the few tens of meters of cryovolcanic residue is not enough ice to completely remove kilometre-sized craters from the surface, when coupled with more extreme relaxation early in Charon's history, the oldest craters are likely to be very eroded and difficult to find.

\section{Fate of Debris: Ejections into the Solar System} \label{sols}

\subsection{Ejections and the Disc}

\begin{figure}
\centering
\includegraphics[scale=.45]{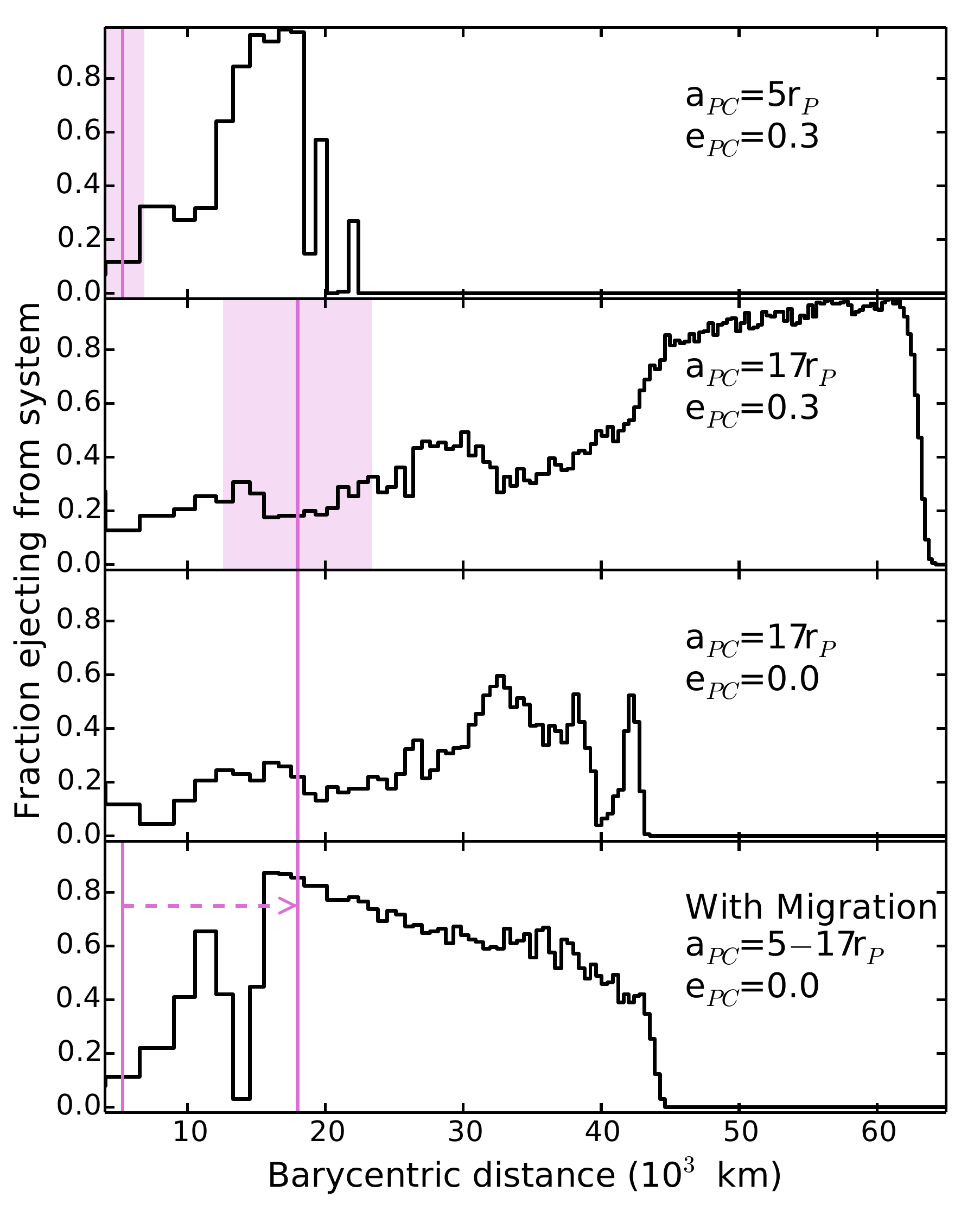}
\caption{  The fraction of disc particles ejected from the \pc\ system per radial bin as a function of barycentric distance.  The figure follows the same style as Figure~\ref{colls}.  Ejections become more common in the outer parts of the disc. In the migration case, the sharp increase in ejections outwards of 12000 km occurs because the outer edge of the instability region sweeps through the disc and perturbs bodies such that they eject instead of collide.  Ejections are nonexistent outside the instability boundary because the disc has not been perturbed.
\label{ej}}
\end{figure}

As the binary instability boundary sweeps out through the disc with Charon's migration, debris is more likely to be ejected from the system than collide with either member of the binary.  We show the ejecting fraction per radial bin in Figure~\ref{ej} for the four \pc\ orbits described in Section~\ref{coll:disc}.  In the three non-migrating orbits, ejections increase with radial distance in the disc.  When migration is included, ejections constitute a high fraction of particle loss from the outer edge of the initial instability boundary to the final instability boundary.  The slight decrease in ejection fraction  as a function of radius occurs because particles are either put on semi-stable eccentric or inclined orbits or have not yet been destructively perturbed (put on an orbit that leads to loss).  We do not include collisional damping of disc particles, which \cite{Walsh2015} show can help stabilise debris, especially in orbits close to the instability boundary.  Collisional damping time-scales are on the order of a few tens to a few hundred years (about an order of magnitude shorter than the instability time-scale), so this could have significant implications for the survival of a ring near the instability boundary.

We record the positions and velocities of all ejected test particles with respect to the \pc\ barycentre when the test particles reach our ejection radius of 0.06 AU, roughly Pluto's Hill radius at 40 AU.  The particles are ejected at 1--15+ times Pluto's escape velocity at the Hill sphere.  Charon's eccentricity is the main factor in the ejection velocity; because encounter velocities are typically lower in the eccentric case, ejection velocities are correspondingly lower (typically by a factor of 2--3). This effect is invariant of Charon's semi-major axis. The majority of ejections in the eccentric case have velocities between 1--5 times escape velocity, while the circular case tends to have ejections with velocities 3--10 times escape velocity. 

\subsection{Debris in the \sols}

We release the ejected test particles from the isolated \pc\ simulation with orbit $a=17$\rp\ and $e=0.3$ into the \sols\ so that we can track the evolution of the \pc\ disc debris.  We simulate three configurations of the modern-day \sols\ (with Pluto at $M=0$, 90, and 180\deg, respectively) and one with the migration model presented in \cite{Malhotra1995}; we orient the \pc\ disc in two ways, labeled ``Aligned with P-C heliocentric orbit'' and ``Misaligned with P-C heliocentric orbit'', as described in Section~\ref{method:sols}.  The fates of these disc-ejected test particles after 1.5 Gyr are shown in Table~\ref{tab:fate}.   The integration including migration retains the largest number of test particles, and the integrations with a misaligned disc retain more test particles in all four initial \sols\ configurations.  The number of collisions is roughly constant throughout; the most common collisions by far are with Jupiter, which account for nearly 50\% of all collisions. Nearly equal numbers of collisions occur with Neptune and Saturn at 10--20\%\ each, while collisions with the Sun and Uranus are rare (less than 10\%\ each in most cases).  Collisions with Pluto are non-existent.

\begin{figure*}
\centering
\includegraphics[scale=.4]{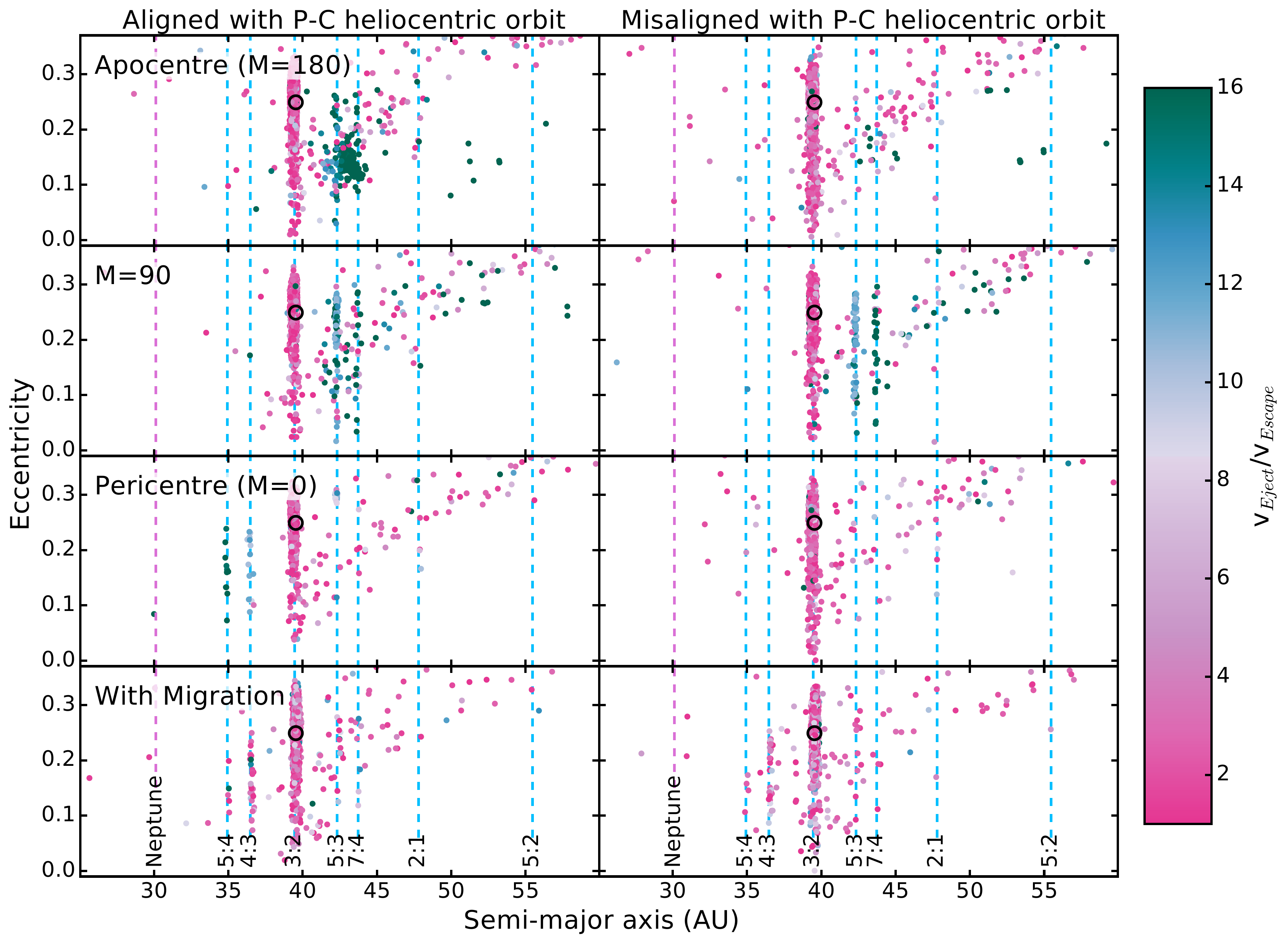}
\caption{ Eccentricity vs. semi-major axis after 1.5 Gyr for wight sets of simulations in which ejected debris from Pluto is released into the \sols. The panels, from top to bottom, show debris initially released when Pluto is at apocentre, $M=90$\deg, pericentre, and in a \sols\ with migration.  The columns show the results for different initial orientations of the \pc\ disc. Colour denotes the initial velocity of the particle (when it leaves Pluto's Hill sphere) relative to the escape velocity from Pluto.  The black open circles show the current location of Pluto.  The dashed line on the left shows the location of Neptune, while the blue dashed lines show the locations of first, second, and third order resonances.  While most of the particles are ejected, 60--80\% of those that remain are trapped in the 3:2 as a population of plutinos.  The majority of other remaining particles, which also tend to be the initially higher velocity particles, are captured into nearby resonances.  
\label{ae}}
\end{figure*}

The semi-major axes and eccentricities of the remaining particles are shown in Figure~\ref{ae}.  In all cases, the majority of the remaining particles populate the 3:2 resonance alongside Pluto.  Nearby resonances can also be populated.  Specifically, we see resonant populations around the 5:4, 4:3, 5:3, and 7:4 resonances with Neptune; some other higher-order resonances may also be populated in between those listed.  We show the fraction of test particles in each resonance in Table~\ref{tab:res}.  In the simulations with a non-migrating \sols, test particles with high initial velocities are more likely to escape the 3:2.  High velocity test particles are lost more frequently in the misaligned disc, leaving a more dominant population in the 3:2. The majority of bodies that end a simulation in a resonance other than the 3:2 are near their initial resonance at the beginning of the simulation due to the additional velocity from ejection modifying the orbital elements.  For instance, a body ejected at Pluto's heliocentric apocentre with 10 times Pluto's escape velocity at the Hill sphere (meaning an addition of 140 m/s to Pluto's orbital velocity of 3.7 km/s) will have a new semi-major axis of 41.4 AU compared with Pluto's semi-major axis of 39.5 AU.  This places the body just interior to the 5:3 resonance.  The occupation of other resonances is more common for release near apocentre  in the aligned simulations because a small velocity change in the particle can cause a larger change in the semi-major axis at apocentre.  We see this trend disappear in the misaligned disc because the high velocity particles are put onto unstable orbits when ejected from the disc. In the migrating case, though, there does not appear to be a preference with initial velocity.  Particles in all five resonances are consistent with libration around the centre of $\phi_{\textrm{res}}=180$\deg.  Those in the 3:2 specifically tend to librate with a higher amplitude than Pluto's $\Delta\phi_{\textrm{res}}=\pm82$\deg: the amplitude of the libration of the resonant angle is closer to 120\deg.

In the eight simulations presented here, a population of debris ejected from the \pc\ disc always remains in the \sols. These bodies would constitute a \pc\ collisional family.  Using the same methodology as presented above to determine the number of craters on Charon's surface, we can estimate the properties of members of the family.  We use the same disc mass and extent as before, but we relax the size constraints to allow the debris to grow to 30 km in radius.  This is about the size of Hydra.  The inner cutoffs of the disc are maintained, as it would take time for debris to coagulate into large sizes observable in the Kuiper belt.  We take ``large'' debris that could be observed by future surveys and constitute a collisional family to be greater than 10 km in radius.  For the \pc\ orbit with  $a=17$\rp\ and $e=0.3$, we calculate that there would be a maximum of 70 objects ejected from the disc larger than 10 km.  For the migrating \pc, a maximum of 200 objects larger than 10 km would be ejected.  Only 7--21\% of these bodies survive to 1.5 Gyr  with either disc orientation. We therefore estimate that there could be 5--15 large KBOs stemming from the Pluto disc that gives the most collisions, while the disc that gives the most ejections seeds the Kuiper belt with 14--42 bodies.

 While we have used a smooth migration model in this work, it is possible that the early history of the \sols\ was more chaotic \citep[e.g., the Nice model from][among others]{Tsiganis2005,Levison2008}. As long as a \sols\ migration model can place Pluto in the correct orbit, our results should be relatively unaffected.  Because the debris ejected from the \pc\ system is very dynamically similar to other Kupier belt populations, there should be similar dynamical evolution between ejected debris and the resonant Kuiper belt objects. 

These members of a Pluto collisional family would be difficult to distinguish from other KBOs.  The particles in our simulations do not have any obvious association  in physical space or orbital angles with Pluto at the end of the simulations (perhaps a slight clustering in both $\omega$, the argument of pericentre, and $\Omega$, the longitude of ascending node).  The most promising method to identify KBOs as members of a Pluto family is through composition; they should have a composition similar to Pluto's icy moons.   Additionally, following the method of \cite{Brown2007} in calculating the velocity dispersion of the collisional family, the plutino members of our remaining debris have a low velocity dispersion of order 100--200 m/s.

\tabcolsep=0.105cm
\begin{table}
\centering
\caption{ Fate of particles in \sols\ integrations: The first column shows the orientation of the debris disc at the start of the simulation, the second column shows the starting position of Pluto, the third shows the number of particles in the sample, and the final three columns show the percentage of particles that remain in the simulation, are ejected, or collide with massive bodies or the Sun.} 
\label{tab:fate}
\begin{tabular}{llcccc}
\hline
&Integration & N$_{\textrm{tot}}$ & Remain & Eject & Collide  \\
&            &  $ \times 10^3$ & \multicolumn{1}{c}{\%} & \multicolumn{1}{c}{\%} & \multicolumn{1}{c}{\%}      \\
\hline
\multirow{ 4}{*}{\makecell[l]{\emph{Aligned}\\\emph{with P-C}\\\emph{heliocentric}\\\emph{orbit}}} & Apocentre ($M=180$) 	& 16 & 13.1 & 85.5 & 1.4 \\
                                                                                                   &$M=90$ 	                & 16 & 8.7  & 89.9 & 1.5 \\
                                                                                                   &Pericentre ($M=0$)   	& 16 & 6.7  & 92.0 & 1.3 \\
                                                                                                   &With Migration 	       & 9.6 & 20.7 & 78.2 & 1.1 \\
\hline
\multirow{ 4}{*}{\makecell[l]{\emph{Misaligned}\\\emph{with P-C}\\\emph{heliocentric}\\\emph{orbit}}} & Apocentre ($M=180$) 	& 16 & 14.4 & 81.2 & 1.4 \\
                                                                                                      &$M=90$ 	                & 16 & 8.9  & 89.4 & 1.7 \\
                                                                                                      &Pericentre ($M=0$)   	& 16 & 11.4 & 87.3 & 1.3 \\
                                                                                                      &With Migration 	     & 10.56 & 17.3 & 81.4 & 1.3 \\
\hline
\end{tabular}
\end{table}

\tabcolsep=0.13cm
\begin{table}
\centering
\caption{Fraction of remaining particles in resonances: The first column shows the orientation of the debris disc at the start of the simulation, the second shows the starting position of Pluto, the third shows the number of particles that remain in the simulations after 1.5 Gyr, and the last five columns show the percentage of remaining particles that fall into the listed first, second, and third order resonances. } 
\label{tab:res}
\begin{tabular}{llcccccc}
\hline
                                                                                                  &Integration & N$_{\textrm{rem}}$ & 3:2 & 5:4 & 4:3 & 5:3 & 7:4 \\
                                                                                                  &            &               & \%  & \%  & \%  & \%  & \%  \\
\hline
\multirow{ 4}{*}{\makecell[l]{\emph{Aligned}\\\emph{with P-C}\\\emph{heliocentric}\\\emph{orbit}}} & Apocentre  & 2112 & 67.1 & 0.1 & 0.1 & 6.3 & 6.3 \\
                                                                                                   &$M=90$      & 1423 & 61.6 & 0.1 & 0.1 & 6.7 & 2.2 \\
                                                                                                   &Pericentre  & 1065 & 60.2 & 1.0 & 1.4 & 3.2 & 0.9 \\
                                                                                                   &With Migration  & 2040 & 79.6 & 0.3 & 1.5 & 1.0 & 0.6 \\
\hline
\multirow{ 4}{*}{\makecell[l]{\emph{Misaligned}\\\emph{with P-C}\\\emph{heliocentric}\\\emph{orbit}}} & Apocentre  & 2793 & 81.2 & 0.1 & 0.2 & 0.5 & 0.7 \\
                                                                                                      & $M=90$     & 1437 & 65.6 & 0.1 & 0.1 & 4.7 & 1.8 \\
                                                                                                      & Pericentre & 1824 & 74.0 & 0.2 & 0.1 & 0.7 & 0.4 \\
                                                                                                      & With Migration & 2259 & 83.3 & 0.3 & 1.6 & 0.9 & 0.3 \\
\hline
\end{tabular}
\end{table}

\section{Conclusions and Discussion} \label{disc}

This work aims to investigate the impact of a debris disc from the Charon-forming giant impact in both the \pc\ system and in the Kuiper belt.  We present \nb\ simulations of the isolated \pc\ binary to follow the fates (collisions and ejections) of debris in the disc, and we also present simulations of the evolution of this debris in the \sols.  We find the following:

\begin{enumerate}
\item The current circumbinary moons, Styx, Nix, Kerberos, and Hydra, did not form in situ if Charon has an eccentric tidal evolution history.  The \cite{Holman1999} instability boundary crosses at least one of the moons' current positions if Charon has $e>0.048$ at its current semi-major axis; many realisations of the Charon-forming impacts from \cite{Canup2005,Canup2011} have the moon forming with eccentricity from 0.1--0.8.  Thus, circumbinary moon formation mechanisms must either invoke a circular tidal evolution for Charon  (or one that leaves Charon on a circular orbit long before it reaches its current semi-major axis) or involve forming the moons after Charon reached its current orbit (through capture, disruption of other bodies, or some other mechanism).
\item  The predominant loss mechanisms in a debris disc around \pc\ are collisions with Charon and ejections.  The amount of clearing is a strong function of Charon's eccentricity.  Collisions are most common from particles that begin close to Charon, while ejections begin to dominate in the outer disc.  Including migration in the simulations causes ejections to increase dramatically because the binary instability boundary interacts with previously unperturbed disc material as the \pc\ orbit expands.  \cite{Walsh2015} find that including collisional evolution in the disc may stabilise material on shorter time-scales, but interactions with the instability boundary will always cause particle loss.
\item Collisions with Charon are most common for a wide orbit, eccentric Charon, such as may have existed near the end of the tidal evolution process. Collisions with Charon are the least common if migration is included. Ejections dominate exterior to Charon's orbit in the wide, eccentric case when Charon undergoes eccentric tidal evolution because the instability boundary is large.  Ejections dominate the majority of the disc in the migration case because of the moving instability boundary. 
\item Assuming a reasonable, realistic (albeit optimistic) disc from \cite{Canup2011} and accounting for a surface solidification time-scale of a few hundred to a few thousand years, we predict hundreds to thousands of craters visible by \emph{New Horizons} on the surface of Charon that stem from the disc and not incident KBO collisions.  It would be difficult to disentangle these populations from size alone as crater-to-impactor size ratios (collisional velocities) are similar.  It is probable that the debris has a different size distribution than KBOs,  in addition to a different average impact velocity, so the presence of two distinct crater populations on the surface of Charon might give insight into the disc.  The apparent lack of small craters on Charon noted by \cite{Singer2016} already has implications on the  extent or composition of a debris disc.
\item Violent tidal heating during the early tidal evolution of Charon may prevent craters from forming on the surface.  If the surface solidifies while tidal heating is still warming the interior, craters should form but may relax.  Radiogenic heating later in the system's history (Gyrs after formation) may contribute to more surface relaxation and/or cryovolcanic resurfacing.  While neither process should cause the oldest craters to disappear, old craters, such as those originating from the debris disc, may appear to be filled in or have indistinct crater walls.  The physical appearance of old craters may help distinguish craters from the disc and craters from KBOs.  
\item About 80--90\% of material ejected into the \sols\ is lost within 1.5 Gyr, regardless of initial Pluto position or inclusion of migration.  The material that remains tends to reside in the 3:2 resonance with Neptune, thereby maintaining a similar orbit to Pluto.  Some material populates nearby resonances, especially the 5:4, 4:3, 5:3, and 7:4 resonances. The material that remains does not show any strong correlation at the end of the simulations with the initial position of Pluto or the other planets.  The objects in resonances, especially in the 3:2, have resonant angles consistent with librating orbits.
\item Using the same methodology that was used to calculate crater numbers and the most optimistic ejection profile, we estimate anywhere from 14--42 icy bodies greater than 10 km in radius could be produced through ejections from the \pc\ disc, forming a ``Pluto disc collisional family.''  Larger, more easily observable members of a Pluto collisional family may originate from the Charon-forming impact itself, such as is seen with the Haumea collisional family.  Members of the collisional family should have similar icy composition to the original disc and a low velocity dispersion. We find no evidence that a collisional family will be disrupted by the migration of the giant planets in the early \sols, nor will it be disrupted through secular or resonant effects over Gyr time-scales. 
\end{enumerate}

The formation of the \pc\ binary and its moons remains both a fascinating and frustrating problem, especially with the enhanced view of the system provided by the \emph{New Horizons} flyby in July 2015.  Through potentially observable tracers such as craters from the debris disc on the surface of Charon or the presence of a Pluto collisional family, we might be able to better constrain the formation and early evolution of this intriguing system.

\section*{Acknowledgements}

Our gratitude to our anonymous referee for insightful comments that improved this manuscript.  Sincere thanks to Alex Parker, Renu Malhotra, Kathryn Volk, Andrew Shannon, and Matija {\'C}uk for helpful discussions pertaining to this work. RAS is supported by the National Science Foundation under Grant No. AST-1410174 and Grant No. DGE-1143953. KMK is supported by the National Science Foundation under Grant No. AST-1410174.  The numerical simulations presented herein were run on the El Gato supercomputer, which is supported by the National Science Foundation under Grant No. 1228509.


\bsp	
\label{lastpage}
\end{document}